\documentclass[%
 reprint,
superscriptaddress,
nofootinbib,
 amsmath,amssymb,
aps,
pra,
]{revtex4-1}

\usepackage{graphicx}
\usepackage{dcolumn}
\usepackage{bm}
\usepackage{hyperref}
\usepackage{amsthm}
\usepackage{bbold}

\begin{document}

\preprint{APS/123-QED}

\title{Rotating wave approximation for quadrupole interaction with high spin}

\author{Wenkui Ding}
\email{wenkuiding@zju.edu.cn}
\affiliation{Zhejiang Institute of Modern Physics and Department of Physics,
Zhejiang University, Hangzhou, Zhejiang 310027, China}

\author{Xiaoguang Wang}
\email{xgwang1208@zju.edu.cn}
\affiliation{Zhejiang Institute of Modern Physics and Department of Physics,
Zhejiang University, Hangzhou, Zhejiang 310027, China}
\affiliation{Graduate School of China Academy of Engineering Physics, Beijing, 100193, China}

\date{\today}

\begin{abstract}
Rotating wave approximation in a quantum spin system driven by a linearly polarized alternating magnetic field with quadrupole interaction presents is investigated in detail in this paper.
The conventional way to employ the rotating wave approximation is to assume the dynamics being restricted in the reduced Hilbert space.
However, when the driving strength is relatively strong or the driving is off resonant, the leakage from the target resonance subspace along with the effects from the counter-rotating terms cannot be neglected anymore.
We propose the rotating wave approximation applied in the full Hilbert space to take the leakage from the target resonance subspace into account.
To include the effects from the counter-rotating terms, we utilize the counterrotating hybridized rotating wave method in the reduced Hilbert space.
The performance of these rotating wave approximation methods is compared by estimating the state fidelity as well as the operator fidelity and the results reveal that different methods may be employed for different practical circumstances.

\end{abstract}

\maketitle


\section{Introduction}
Rotating wave approximation (RWA) plays an important role in many fields of quantum physics, such as magnetic resonance~\cite{abragam1961principles,slichter2013principles,mehring2012principles}, quantum optics~\cite{allen1987optical,scully1997quantum}, etc.
Particularly, along with the RWA, the rotating reference frame has a wide application in the theory of magnetic resonance~\cite{rabi1954use}.
However, the feasibility of the standard RWA requires some conditions, such as the driving field is on resonance or near resonance, the strength of the driving is weak, etc.
When these conditions are violated~\cite{shirley1965solution,ashhab2007two}, the counter-rotating terms will lead to many significant effects, such as the Bloch-Siegert shifts~\cite{bloch1940magnetic}, the coherent destruction of tunneling~\cite{grossmann1991coherent,barata2000strong}, and so on. 
Many works have dedicated to improve the performance of the rotating wave approximation~\cite{lv2012effects,wu2007strong,irish2005dynamics,irish2007generalized,hausinger2010dissipative}, however, these works mainly focused on  the two-level system or systems with su(2) algebra.

The knowledge on the dynamics of a quantum spin system driven by a linearly polarized ac field is important for many applications, such as the high-fidelity quantum control or state preparation~\cite{wiseman2009quantum}, high-sensitivity and high-precision quantum sensing~\cite{degen2017quantum}, reliable quantum computing~\cite{nielsen2002quantum}, and so on.
Since the quantum spin system driven by a linearly polarized ac magnetic field is the physical model of many realistic quantum systems, it has been widely studied in many fields of quantum physics~\cite{allen1987optical,abragam1961principles,vandersypen2005nmr}.
Specifically, the high-spin quantum system (with spin $I>1/2$) reveals some advantages in quantum sensing~\cite{lee2015vector}.
However, the quadrupole interaction always presents in a high-spin quantum system~\cite{slichter2013principles}, and the standard rotating wave approximation needs to be modified before applying.
In this paper, we carefully investigate the rotating wave approximation in a high-spin quantum system with quadrupole interaction presents.

For a quantum spin system with quadrupole interaction driven by a linearly polarized ac magnetic field, the Hamiltonian can be written as~\cite{slichter2013principles,abragam1961principles,marcus2013nitrogen,mizuochi2002continuous}
\begin{equation}
\label{eq:original_Hamiltonian}
H(t)=QI_z^2+B_0I_z+B_1\cos(\omega t)I_x,
\end{equation}
where the first term corresponds to the quadrupole interaction\footnote{For systems like nitrogen-vacancy color center in diamond, this interaction is usually called zero-field splitting, which results from many-body interactions.} for spin $I>1/2$, and $Q$ is the quadrupole coupling strength.
The second term is the Zeeman energy, where $B_0$ is the strength of the static magnetic field, which is usually applied to lift the degeneracy induced by the quadrupole interaction.
The third term corresponds to the driving of the quantum spin system by a linearly polarized ac magnetic field, where $B_1$ and $\omega$ are the amplitude and frequency of the oscillating magnetic field, respectively.
Throughout this paper we set $\hbar=1$ and assume the gyromagnetic ratio of the spin $\gamma_{I}=1$.
Usually, the interaction of the quantum spin system with an alternating magnetic field can be approximated by the interaction of the quantum spin system with a rotating magnetic field.
This is done by decomposing the interaction of the spin with the alternating magnetic field into the co-rotating term and the counter-rotating term, then the counter-rotating term is dropped due to its fast oscillation in the interaction picture. 
This is the well-known standard rotating wave approximation.

Firstly, we consider the situation that Zeeman splitting is much larger than the quadrupole coupling, namely, $B_0\gg Q$.
The effect of the quadrupole coupling is to induce deviations of the resonance frequency between different Zeeman levels.
In fact, the standard rotating wave approximation can be directly applied for this situation.
We illustrate the application of the RWA in this situation by transforming the original Hamiltonian [Eq.~(\ref{eq:original_Hamiltonian})] into the interaction picture (set by $H_0=B_0I_z+QI_z^2$),
\begin{equation}
\label{eq:inter_pic_H}
\begin{aligned}
H_I(t)=&e^{iH_0t}H(t)e^{-iH_0t}-H_0\\
=&\frac{B_1}{2}e^{i(B_0I_z+QI_z^2-\omega I_z)t}I_xe^{-i(B_0I_z+QI_z^2-\omega I_z)t}\\
&+\frac{B_1}{2}e^{i(B_0I_z+QI_z^2+\omega I_z)t}I_xe^{-i(B_0I_z+QI_z^2+\omega I_z)t}.
\end{aligned}
\end{equation}
When the driving is near resonance $\omega\sim B_0$, and since $B_0\gg Q$, we only keep the first (slowly oscillating) term for the RWA.
We then transform back to the Schr\"{o}dinger picture and the Hamiltonian after applying the rotating wave approximation now becomes
\begin{equation}
H(t)\approx QI_z^2+B_0I_z+\frac{B_1}{2}e^{-i\omega I_zt}I_xe^{i\omega I_zt}.
\end{equation}
Fortunately, we can exactly solve the dynamics of this Hamiltonian in the rotating reference frame by applying the unitary transformation $U(t)=e^{-i\omega I_zt}$.
The effective Hamiltonian in the rotating frame then becomes time-independent,
\begin{equation}
H_{\text{eff}}=(B_0-\omega)I_z+QI_z^2+\frac{B_1}{2}I_x.
\end{equation}
Finally, the evolution operator corresponding to Hamiltonian Eq.~(\ref{eq:original_Hamiltonian}) after employing the RWA for the situation that Zeeman interaction dominates can be written explicitly as
\begin{equation}
\mathcal{U}(t)=e^{-i\omega I_z t}e^{-iH_{\text{eff}}t}.
\end{equation}


However, when quadrupole interaction dominates, namely, $Q\gg B_0$, the situation changes dramatically and the above procedure to directly apply the standard rotating wave approximation cannot be employed anymore.
To clearly elaborate this fact, we first illustrate the difficulty in directly applying the rotating wave approximation for a spin system with quadrupole coupling only (temporarily neglect the Zeeman interaction).
The Hamiltonian then becomes $H^\prime=QI_z^2$, and the initial state of the spin system is prepared in the eigenstate of $I_x$,
\begin{equation}
|\psi(0)\rangle=e^{-i\frac{\pi}{2}I_y}|I,I\rangle=\sum_{M=-I}^Ic_M|I,M\rangle,
\end{equation}
where $|I,M\rangle$ is the eigenstate of $I_z$, and $M$ is the eigenvalue of $I_z$. 
The coefficients $c_M$'s satisfy the following relation:
\begin{equation}
\begin{aligned}
c_{M-1}=&c_M\frac{2I}{\sqrt{(I-M+1)(I+M)}}\\
&-c_{M+1}\frac{\sqrt{(I+M+1)(I-M)}}{\sqrt{(I-M+1)(I+M)}}.
\end{aligned}
\end{equation}
The time evolution of the spin vector, which is defined as $\vec{\mathbf{V}}(t)=\langle I_x(t)\rangle\hat{\mathbf{x}}+\langle I_y(t)\rangle\hat{\mathbf{y}}+\langle I_z(t)\rangle\hat{\mathbf{z}}$, where $\langle I_{i=x,y,z}(t)\rangle=\langle \psi(0)|I_i(t)|\psi(0)\rangle$ is the expectation value of the spin operator, can be straightforwardly calculated using the Heisenberg equation, $dI_i(t)/dt=i[H^\prime, I_i(t)]$. 
The calculated results are
\begin{equation}
\begin{aligned}
&\langle I_x(t)\rangle=\\
&\sum_{M=-I}^{I-1}\sqrt{(I-M)(I+M+1)}c_Mc_{M+1}\cos[(2M+1)Qt],\\
&\langle I_y(t)\rangle=\langle I_z(t)\rangle=0,
\end{aligned}
\end{equation}
and after some simplifications, we obtain
\begin{equation}
\vec{\mathbf{V}}(t)=I[\cos(Qt)]^{2I-1}\ \hat{\mathbf{x}},
\end{equation}
which can be decomposed into clockwise rotating terms and counterclockwise rotating terms.
Specifically, when spin $I$ is integer,
\begin{equation}
\label{eq:spin_vector_integer}
\begin{aligned}
\vec{\mathbf{V}}(t)_{\text{integer}}=\frac{I}{2^{2I-2}}\sum_{k=0}^{I-1}C_{2I-1}^k[\vec{\mathbf{V}}_{R,k}(t)+\vec{\mathbf{V}}_{L,k}(t)],
\end{aligned}
\end{equation}
where $C_n^k$ is the binomial coefficient, and
\begin{eqnarray}
\vec{\mathbf{V}}_{R,k}(t)=\cos{\omega_k t}\ \hat{\mathbf{x}}+\sin{\omega_k t}\ \hat{\mathbf{y}},\\
\vec{\mathbf{V}}_{L,k}(t)=\cos{\omega_k t}\ \hat{\mathbf{x}}-\sin{\omega_k t}\ \hat{\mathbf{y}},
\end{eqnarray}
with $\omega_k=(2I-1-2k)Q$.
Similarly, when spin $I$ is half-integer,
\begin{equation}
\label{eq:spin_vector_half}
\begin{aligned}
\vec{\mathbf{V}}(t)_{\text{half}}=&\frac{I}{2^{2I-1}}C_{2I-1}^{I-1/2}\\
&+\frac{I}{2^{2I-2}}\sum_{k=0}^{I-3/2}C_{2I-1}^k[\vec{\mathbf{V}}_{R,k}(t)+\vec{\mathbf{V}}_{L,k}(t)].
\end{aligned}
\end{equation}

These results show that, the time evolution of the spin vector under the quadrupole interaction is different from the time evolution under Zeeman interaction.
For the latter, the spin vector only rotates in one direction, where we can easily define the co-rotating component and the counter-rotating component of the ac magnetic field.
On the other hand, for the case with quadrupole interaction, the evolving spin vector is decomposed into many clockwise rotating vectors and counterclockwise rotating vectors, along with different rotating frequencies.
To make a summary, the unitary operator, $\exp(-iQI_z^2t)=\exp[-i(QI_zt)I_z]$, can be regarded as the rotation around the $z$ axis, but the rotation frequency and direction are dependent on $I_z$ at the same time.
It is this characteristic that results in the difficulty to directly employ the standard rotating wave approximation.
Besides, from Eq.~(\ref{eq:spin_vector_integer}) and Eq.~(\ref{eq:spin_vector_half}), we can see that the case with integer spin shows higher symmetry property than the case with half-integer spin.
In the following sections, we will first focus on the investigation on the case with integer spin and then generalize the results to the spin half-integer case.

In order to provide a complete analysis on the rotating wave approximation for the high-spin system, we organize our paper as follows.
In Sec.~\ref{sc:rwa_reduce}, we discuss the conventional method usually used in the multi-level quantum system, namely, reducing the Hilbert space, to apply the rotating wave approximation.
In Sec.~\ref{sc:rwa_full}, we propose our procedure to apply the rotating wave approximation in the full Hilbert space, where the effect of the leakage from the target resonance subspace is taken into account.
In Sec.~\ref{sc:chrw}, we investigate the counterrotating hybridized rotating wave method in the reduced Hilbert space for the high-spin system with quadrupole interaction, where the effect from the counter-rotating terms is taken into account.
In Sec.~\ref{sc:comparison}, we calculate the state fidelity and the operator fidelity by exactly solving the dynamics, and compare the performance of the RWA procedures proposed in this paper.
Finally, in Sec.~\ref{sc:discussion}, we discuss the problems that need to be investigated in the future and the possible applications of our results.

\section{Rotating wave approximation for the reduced Hilbert space\label{sc:rwa_reduce}}

In this section, we will first consider the case with integer spin $I$, and discuss the case with half-integer spin in the end.
Firstly, we return to the original Hamiltonian Eq.~(\ref{eq:original_Hamiltonian}), where the resonance frequency between levels $|I,M\rangle$ and $|I,M-1\rangle$ is
\begin{equation}
\omega_M=(2M-1)Q+B_0.
\end{equation} 
Since we now focus on the situation that quadrupole interaction dominates, namely, $Q\gg B_0$, the resonance frequency $\omega_M\sim |(2M-1)Q|$, which coincides with the resonance frequency between levels $|I,-M+1\rangle$ and $|I,-M\rangle$.
Usually, when the frequency of the applied ac magnetic field is near resonance ($\omega \sim \omega_M$) and the driving strength is relatively weak ($B_1\ll \omega$), we can reduce the Hilbert space to employ the standard rotating wave approximation.
For instance, when the frequency of the ac magnetic field $\omega\sim |(2M-1)Q|$, significant transitions occur between states $|I,M\rangle$ and $|I,M-1\rangle$, along with between states $|I,-M+1\rangle$ and $|I,-M\rangle$.
When the detuning of the resonance is far from the nearest level splitting, namely, $\omega-|(2M-1)Q|\ll |(2M-3)Q|$, the dynamics can be restricted in the subspace spanned by the basis states, $|I,M\rangle$, $|I,M-1\rangle$, $|I,-M+1\rangle$ and $|I,-M\rangle$; since transitions between other states get greatly suppressed.
Moreover, when $|M|>1$, because the dynamics in the $|I,M\rangle$, $|I,M-1\rangle$ subspace is decoupled from the dynamics in the $|I,-M+1\rangle$, $|I,-M\rangle$ subspace, the restricted Hilbert space can be further treated as two decoupled two-level systems.
On the other hand, when $|M|=1$, the reduced Hilbert space is spanned by $|I,M=1\rangle$, $|I,M=0\rangle$ and $|I,M=-1\rangle$, which is an effective three-level system.

\subsection{Dynamics in the reduced two-level systems}
Firstly, we consider the case that the frequency of the applied ac magnetic field $\omega\sim (2M-1)Q$ with $|M|>1$.
For this case, significant transitions of the system primarily take place in two decoupled two-level systems, thus the original Hamiltonian can be approximately separated into three subspaces,
\begin{equation}
H(t)\approx H_1(t)+H_2(t)+H_n(t),
\end{equation}
where $H_1(t)$ corresponds to the Hamiltonian projected to the $|I,M\rangle$, $|I,M-1\rangle$ subspace; $H_2(t)$ corresponds to the Hamiltonian projected to the $|I,-M+1\rangle$, $|I,-M\rangle$ subspace; $H_n(t)$ corresponds to the Hamiltonian projected to the rest of the Hilbert space.
It is clear that $[H_1(t),H_2(t)]=0$, $[H_1(t),H_n(t)]=0$ and $[H_2(t),H_n(t)]=0$.
After expanding the original Hamiltonian in the $|I,M\rangle$, $|I,M-1\rangle$ subspace, we obtain
\begin{equation*}
\begin{aligned}
H_1(t)=&(QM^2+B_0M-\frac{\omega_0}{2})\mathbb{1}_2+\frac{\omega_0}{2}\hat{\sigma}_z\\
&+\frac{B_1}{2}\sqrt{(I+M)(I-M+1)}\cos(\omega t)\hat{\sigma}_x,
\end{aligned}
\end{equation*}
where $\hat{\sigma}_{i=x,y,z}$ are the Pauli matrices, $\mathbb{1}_2$ is the $2\times 2$ identity matrix, $\omega_0\equiv\omega_M=(2M-1)Q+B_0$ is the resonance frequency between levels $|I,M\rangle$ and $|I,M-1\rangle$.
The first constant term in $H_1(t)$ can be neglected and we can rewrite the Hamiltonian in the form of a quantized spin-1/2 in an alternating magnetic field,
\begin{equation}
\label{eq:two_level}
H_1(t)=\frac{\omega_0}{2}\hat{\sigma}_z+B_1^\prime \cos(\omega t)\hat{\sigma}_x,
\end{equation}
where $B_1^\prime=\frac{B_1}{2}\sqrt{(I+M)(I-M+1)}$ is the modified amplitude of the alternating magnetic field.
The standard rotating wave approximation procedure can now be applied for this subsystem as described previously.
After dropping the fast oscillating term in the interaction picture, the Hamiltonian in the Schr\"{o}dinger picture becomes
\begin{equation}
H_1(t)\approx \frac{\omega_0}{2}\hat{\sigma}_z+\frac{B_1^\prime}{2} e^{-i\omega \hat{\sigma}_z t/2}\hat{\sigma}_x e^{i\omega \hat{\sigma}_z t/2}.
\end{equation}
We then utilize the unitary transformation $U(t)=e^{-i\omega \hat{\sigma}_z t/2}$ to obtain the time-independent effective Hamiltonian in the rotating frame,
\begin{equation}
H_{\text{eff}}=\frac{\omega_0-\omega}{2}\hat{\sigma}_z+\frac{B_1^{\prime}}{2}\hat{\sigma}_x,
\end{equation}
and the evolution matrix corresponding to $H_1(t)$ can be calculated explicitly as
\begin{align*}
\mathcal{U}_1(t)&=e^{-i\omega \hat{\sigma}_z t/2}e^{-iH_{\text{eff}}t}\\
&=\tau_0\mathbb{1}_2-i\frac{\tau_x}{2}\hat{\sigma}_x-i\frac{\tau_y}{2}\hat{\sigma}_y-i\frac{\tau_z}{2}\hat{\sigma}_z,
\end{align*}
where
\begin{align*}
\tau_0&=\cos{\frac{\omega t}{2}}\cos(\Omega t)-\frac{\Delta}{2\Omega}\sin{\frac{\omega t}{2}}\sin(\Omega t),\\
\tau_x&=B_1^{\prime}\frac{\sin(\Omega t)}{\Omega}\cos{\frac{\omega t}{2}},\\
\tau_y&=B_1^{\prime}\frac{\sin(\Omega t)}{\Omega}\sin{\frac{\omega t}{2}},\\
\tau_z&=B_1^{\prime}\frac{\Delta\sin(\Omega t)}{\Omega}\cos{\frac{\omega t}{2}}+2\cos(\Omega t)\sin{\frac{\omega t}{2}},\\
\end{align*}
with $\Delta=\omega_0-\omega$ and $\Omega=\sqrt{\Delta^2+{B_1^\prime}^2}/2$.

The evolution matrix $\mathcal{U}_2(t)$ corresponding to $H_2(t)$ in the $|I,-M+1\rangle$, $|I,-M\rangle$ subspace can be obtained via the same procedure as described above,
\begin{equation}
\mathcal{U}_2(t)=e^{i\omega \hat{\sigma}_z t/2}e^{-iH_{\text{eff}}t},
\end{equation}
along with the change of the resonance frequency $\omega_0=(2M-1)Q-B_0$ in $H_{\text{eff}}$.
Eventually, the overall evolution matrix for the whole Hilbert space can be represented as the direct sum of the evolution matrices in their respective subspace,
\begin{equation}
\label{eq:Ut_two_level}
\begin{aligned}
\mathcal{U}(t)=&\Big[\oplus_{M^\prime=I}^{M+1}e^{-i(Q{M^\prime}^2+B_0M^\prime)t}\Big]\oplus \mathcal{U}_1(t)\\
&\Big[\oplus_{M^\prime=M-1}^{-M+1}e^{-i(Q{M^\prime}^2+B_0M^\prime)t}\Big]\oplus\mathcal{U}_2(t)\\
&\Big[\oplus_{M^\prime=-M-2}^{-I}e^{-i(Q{M^\prime}^2+B_0M^\prime)t}\Big].
\end{aligned}
\end{equation}

\subsection{Dynamics in the reduced three-level system}
Next, we consider the situation that the frequency of the ac magnetic field $\omega\sim Q$, namely, significant transitions occur among levels $|I,M=1\rangle$, $|I,M=0\rangle$ and $|I,M=-1\rangle$.
Similar to previous discussions, when the detuning of the resonance is much smaller than the energy splitting to the nearest level, namely, $|\omega-Q|\ll 3Q$, the dynamics can be restricted to the subspace spanned by these three levels. 
The reduced Hamiltonian in this subspace then becomes an effective Hamiltonian for a three-level system,
\begin{equation}
\label{eq:reduced_spin_1}
H(t)=QS_z^2+B_0S_z+\sqrt{\frac{I(I+1)}{2}}B_1\cos{(\omega t)}S_x,
\end{equation}
where $S_{x,y,z}$ are spin operators for spin $S=1$.
It can be checked that, for spin $S=1$, we have the relation~\cite{ajoy2012stable} that
\begin{equation}
\label{eq:spin1_specific}
2\cos(\omega t)S_x=e^{-i\omega  S_z^2 t}S_xe^{i\omega S_z^2 t}+e^{i\omega S_z^2 t}S_xe^{-i\omega S_z^2 t},
\end{equation}
thus the Hamiltonian can be written as
\begin{equation}
\begin{aligned}
H(t)=&QS_z^2+B_0S_z\\
&+\frac{B_1^\prime}{2}(e^{-i\omega  S_z^2 t}S_xe^{i\omega S_z^2 t}+e^{i\omega S_z^2 t}S_xe^{-i\omega S_z^2 t}),
\end{aligned}
\end{equation}
where we have denoted $B_1^\prime=\sqrt{\frac{I(I+1)}{2}}B_1$.

In order to employ the rotating wave approximation, we first transform the Hamiltonian to the interaction picture by choosing $H_0=QS_z^2+B_0S_z$,
\begin{align*}
H_I(t)=&e^{iH_0t}H(t)e^{-iH_0t}-H_0\\
=&\frac{B_1^\prime}{2}\Big[e^{iB_0S_zt}e^{i(Q-w)S_z^2t}S_xe^{-i(Q-w)S_z^2t}e^{-iB_0S_zt}\\
&+e^{iB_0S_zt}e^{i(Q+w)S_z^2t}S_xe^{-i(Q+w)S_z^2t}e^{-iB_0S_zt}\Big].
\end{align*}
For the rotating wave approximation, we only keep the slowly oscillating term and drop the rapidly oscillating term (here $Q-w\ll Q+w$),
\begin{equation}
H_I(t)\approx\frac{B_1^\prime}{2}e^{iB_0S_zt}e^{i(Q-w)S_z^2t}S_xe^{-i(Q-w)S_z^2t}e^{-iB_0S_zt}.
\end{equation}
Transforming back to the Schr\"{o}dinger picture, the Hamiltonian after applying the rotating wave approximation then becomes
\begin{equation}
H(t)\approx QS_z^2+B_0S_z+\frac{B_1^\prime}{2}e^{-iwS_z^2t}S_xe^{iwS_z^2t}.
\end{equation}
Next, we transform to the rotating reference frame by applying the unitary transformation $U(t)=e^{-i\omega S_z^2 t}$, to make the Hamiltonian time-independent.
In this rotating frame, the effective Hamiltonian becomes
\begin{equation}
\begin{aligned}
H_{\text{eff}}&=U^\dagger(t)H(t)U(t)-iU^\dagger(t)\dot{U}(t)\\
&=(Q-\omega)S_z^2+B_0S_z+\frac{B_1^\prime}{2}S_x.
\end{aligned}
\end{equation}
The evolution matrix for this subsystem now becomes $\mathcal{U}_3(t)=e^{-i\omega S_z^2t}e^{-iH_{\text{eff}}t}$, which can be calculated explicitly as well.
In order to utilize the analytic result for any SU(3) group element generated by a traceless $3\times 3$ Hermitian matrix in Ref.~\cite{thomas2015elementary}, we need to modify and renormalize the Hamiltonian,
\begin{equation}
\mathcal{H}=\sqrt{\frac{2}{u}}\Big[H_{\text{eff}}-\frac{2}{3}(Q-\omega)\mathbb{1}_3\Big],
\end{equation}
where $\mathbb{1}_3$ is the $3\times 3$ unity matrix and $u=2B_0^2+(B_1^\prime)^2/2+2(Q-\omega)^2/3$ is the normalization factor.
Now, the evolution operator can be written explicitly as
\begin{equation}
\label{eq:Ut_explicit_spin1}
\begin{aligned}
\mathcal{U}_3(t)=&\sum_{k=0,1,2}\mathcal{R}\Big[\mathcal{H}^2+\frac{2\sin(\alpha+\frac{2\pi k}{3})}{\sqrt{3}}\mathcal{H}\\
&-\frac{1+2\cos(2\alpha+\frac{4\pi k}{3})}{3}\mathbb{1}_3\Big]\frac{\exp[it\sqrt{\frac{2u}{3}}\sin(\alpha+\frac{2\pi k}{3})]}{1-2\cos(2\alpha+\frac{4\pi k}{3})},
\end{aligned}
\end{equation}
where
\begin{equation*}
\begin{aligned}
\mathcal{R}&=(e^{-i\omega t}-1)S_z^2+\mathbb{1}_3,\\
\alpha&=\frac{1}{3}[\arccos(\frac{3\sqrt{3}}{2}\det \mathcal{H})-\frac{\pi}{2}].
\end{aligned}
\end{equation*}
Eventually, the overall evolution matrix for the whole Hilbert space can be written as the direct sum of the evolution matrices in their respective subspace,
\begin{equation}
\label{eq:spin1_reduced}
\begin{aligned}
\mathcal{U}(t)=&\Big[\oplus_{M=I}^{2}e^{-i(QM^2+B_0M)t}\Big]\\
&\oplus \mathcal{U}_3(t) \Big[\oplus_{M=-2}^{-I}e^{-i(QM^2+B_0M)t}\Big].
\end{aligned}
\end{equation}

In the end, we briefly discuss the case with half-integer spin $I$.
When $\omega\sim (2M-1)Q$ with $|M|>\frac{1}{2}$, the dynamics is approximately restricted in two decoupled two-level systems, just the same as the spin integer case.
On the other hand, when $\omega\sim Q$, the dynamics is approximately restricted in a two-level system spanned by $|I,M=\frac{1}{2}\rangle$ and $|I,M=-\frac{1}{2}\rangle$, and the evolution operator can be obtained similarly by procedures as introduced previously.

\section{Rotating wave approximation for the full Hilbert space\label{sc:rwa_full}}

For the rotating wave approximation applied in the reduced Hilbert space, transitions from the reduced subspace to levels outside the reduced subspace, or transitions between levels outside the reduced subspace, are neglected.
However, these transitions cannot be neglected when the resonance detuning is large or the driving strength is strong. 
We now consider the procedure to employ the rotating wave approximation in the full Hilbert space.
In this section, we also first consider the case with integer spin and then discuss the case with half-integer spin.
The difficulty to apply the RWA in the full Hilbert space lies in the fact that, for the high-spin system, namely, $I>1$, the relation in Eq.~(\ref{eq:spin1_specific}) does not exist, so we have to find a modified procedure to employ the RWA.

\subsection{RWA in the full Hilbert space for integer spin}
The matrix form of Eq.~(\ref{eq:inter_pic_H}) in the basis spanned by $|I,M\rangle$'s can be written explicitly as
\begin{equation}
\begin{aligned}
&H_I(t)=\\
&\sum_{M=-I+1}^IB_1^\prime\Big [(e^{-i(\omega_M-\omega)t}+e^{-i(\omega_M+\omega)t})|I,M-1\rangle\langle I,M|\\
&+(e^{i(\omega_M-\omega)t}+e^{i(\omega_M+\omega)t})|I,M\rangle\langle I,M-1|\Big ],
\end{aligned}
\end{equation}
where $B_1^\prime=\frac{B_1}{4}\sqrt{(I+M)(I-M+1)}$ and $\omega_M=(2M-1)Q+B_0$.
For the rotating wave approximation, we drop these rapidly oscillating terms, namely, the terms with the exponent $\propto (nQ+\omega)$.
After dropping these terms, the Hamiltonian in the interaction picture becomes
\begin{equation}
\begin{aligned}
H_I(t)\approx& \sum_{M=1}^IB_1^\prime\Big [e^{-i(\omega_M-\omega)t}|I,M-1\rangle\langle I,M|\\&+e^{i(\omega_M-\omega)t}|I,M\rangle\langle I,M-1|\Big ]\\
&+\sum_{M=-I+1}^0B_1^\prime\Big [e^{-i(\omega_M+\omega)t}|I,M-1\rangle\langle I,M|\\&+e^{i(\omega_M+\omega)t}|I,M\rangle\langle I,M-1|\Big ].
\end{aligned}
\end{equation}

On the other hand, the above Hamiltonian after applying the rotating wave approximation can also be obtained by introducing a specific operator
\begin{equation}
I_a=\sum_{M=1}^I|I,M\rangle\langle I,M|-\sum_{M^\prime=-I}^{-1}|I,M^\prime\rangle\langle I,M^\prime|.
\end{equation}
This operator is one of the key results obtained in this paper. 
For spin $I=1$, it happened to be $I_a=I_z$, in accordance with previous discussions.
We can check that, the original Hamiltonian Eq.~(\ref{eq:original_Hamiltonian}) after applying the rotating wave approximation in the full Hilbert space actually becomes
\begin{equation}
\label{eq:original}
H(t)\approx QI_z^2+B_0I_z+\frac{B_1}{2}e^{-i\omega tI_zI_a}I_xe^{i\omega tI_zI_a}.
\end{equation}
Accordingly, we can transform this Hamiltonian to the rotating reference frame by applying the unitary transformation
\begin{equation}
U(t)=e^{-i\omega tI_zI_a}.
\end{equation}
The effective Hamiltonian in this rotating frame now becomes time-independent,
\begin{equation}
\begin{aligned}
H_{\text{eff}}&=U^\dagger(t)H(t)U(t)-iU^\dagger(t)\dot{U}(t)\\
&=QI_z^2-\omega I_zI_a+B_0I_z+\frac{B_1}{2}I_x,
\end{aligned}
\end{equation}
where we have used the fact that $[I_a,I_z]=0$.
Finally, the evolution operator for the case with integer spin in the full Hilbert space can be written as
\begin{equation}
\label{eq:Ut_full}
\mathcal{U}(t)=e^{-i\omega tI_zI_a}e^{-iH_{\text{eff}}t}.
\end{equation}

\subsection{RWA in the full Hilbert space for half-integer spin}
Next, we consider the case with half-integer spin.
For the spin angular momentum $\mathbf{J}$ with half-integer quantum number $J$, we can regard it as the addition of two angular momentum,
\begin{equation}
\mathbf{J}=\mathbf{I}+\mathbf{S},
\end{equation}
where $\mathbf{I}$ is the operator of spin angular momentum with integer quantum number $I$ and $\mathbf{S}$ is the spin operator of spin $S=1/2$.
Now, the Hamiltonian for spin $J$ can be rewritten as follows:
\begin{equation}
\begin{aligned}
H(t)=&QJ_z^2+B_0J_z+B_1\cos(\omega t)J_x\\
=&Q(I_z+S_z)^2+B_0(I_z+S_z)+B_1\cos(\omega t)(I_x+S_x)\\
=&QI_z^2+B_0I_z+B_1\cos(\omega t)I_x+QS_z^2\\
&+B_0S_z+B_1\cos(\omega t)S_x+2QI_zS_z.
\end{aligned}
\end{equation}
Since $QS_z^2$ is a constant term, we can neglect it from the Hamiltonian.
Following the procedure introduced above to apply the rotating wave approximation for integer spin $I$ in the full Hilbert space, we first apply the unitary transformation $U_1(t)=\exp({-i\omega tI_zI_a})$, then the effective Hamiltonian after applying this transformation becomes
\begin{equation}
\begin{aligned}
H_{\text{eff}}=&QI_z^2-\omega I_zI_a+B_0I_z+\frac{B_1}{2}I_x\\
&+B_0S_z+B_1\cos(\omega t)S_x+2QI_zS_z.
\end{aligned}
\end{equation}
Next, we apply the second unitary transformation $U_2(t)=\exp({-i\omega tS_z})$ and the effective Hamiltonian after applying this transformation becomes
\begin{equation}
\begin{aligned}
H_{\text{eff}}^\prime=&QI_z^2-\omega I_zI_a+B_0I_z+\frac{B_1}{2}I_x\\
&+(B_0-\omega)S_z+\frac{B_1}{2}S_x+2QI_zS_z.
\end{aligned}
\end{equation}
The evolution operator for the case with half-integer spin in the full Hilbert space can now be written as
\begin{equation}
\label{eq:Ut_full_half}
\mathcal{U}(t)=\exp(-i\omega tI_zI_a)\exp(-i\omega tS_z)\exp(-iH_{\text{eff}}^\prime t).
\end{equation}
It is essential to note that the calculation of the dynamics is usually done in the basis of $|J,M_J\rangle$ and we need to expand Eq.~(\ref{eq:Ut_full_half}) in this basis, which needs the following relation between the basis states~\cite{landau2013quantum},
\begin{equation}
\begin{aligned}
|J=I+\frac{1}{2},M_J\rangle=&\sqrt{\frac{J+M_J}{2J}}|S=\frac{1}{2},M_S=\frac{1}{2}\rangle\\
&\otimes|I,M_I=M_J-\frac{1}{2}\rangle\\
&+\sqrt{\frac{J-M_J}{2J}}|S=\frac{1}{2},M_S=-\frac{1}{2}\rangle\\
&\otimes|I,M_I=M_J+\frac{1}{2}\rangle,
\end{aligned}
\end{equation}
where $|I,M_I\rangle$ is the eigenstate of $I_z$ and $|S,M_S\rangle$ is the eigenstate of $S_z$.

In fact, the procedure introduced above can also be employed to calculate the evolution operator for the spin integer case, instead of using Eq.~(\ref{eq:Ut_full}) that we provided.
However, compared to our simple and compact solution, the iteration method utilizing the addition of angular momentum is complicated and tedious, while the calculated evolution operator is rather intricate than the form of Eq.~(\ref{eq:Ut_full}).

\subsection{Comparison with RWA in the reduced Hilbert space}

\begin{figure}
\includegraphics[width=0.5\textwidth]{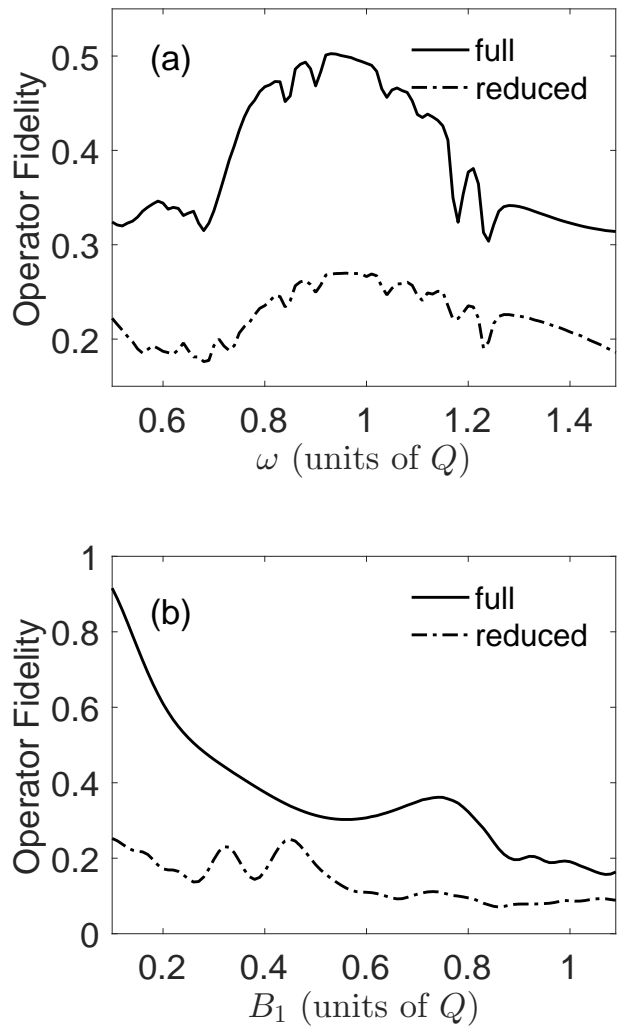}
\centering
\caption{\label{fig:compare_full_reduced}Comparison of the averaged operator fidelity after applying the rotating wave approximation for spin $I=3$ in the full Hilbert space (solid lines) or in the reduced Hilbert space (dash-dotted lines). The initial state of the system is $|\psi(0)\rangle=|I=3,M=0\rangle$, with the ac magnetic field applied to induce transitions between $|I=3,M=\pm 1\rangle$ and $|I=3,M=0\rangle$. (a) Operator fidelity as a function of the frequency of the ac magnetic field, with the fixed amplitude of the ac magnetic field $B_1=0.5Q$ and the static magnetic field $B_0=0.05Q$. (b) Operator fidelity as a function of the strength of the ac magnetic field, with the fixed detuing ($\omega=1.5Q$).}
\end{figure}

We compare the performance of the RWA applied in the full Hilbert space with the RWA applied in the reduced Hilbert space by calculating the operator fidelity which is defined in Eq.~(\ref{eq:operator_fidelity}) (see more details in Sec.\ref{sc:comparison}).
Basically, we investigate the effect of the detuning and the driving strength on the performance of the rotating wave approximation.
The calculated result is presented in Fig.~\ref{fig:compare_full_reduced}, where the value of the operator fidelity is averaged over twenty $\pi$ rotation cycles, namely, the evolution time is $20T_{\pi}$ [see Eq.~(\ref{eq:t_pi}) for the exact definition of $T_{\pi}$].
In Fig.~\ref{fig:compare_full_reduced}(a), the driving strength of the ac magnetic field is relatively strong, with a fixed value $B_1=0.5Q$, while the frequency of the ac magnetic field is varied around the on resonant value $\omega\sim Q$.
It is clear that, for both cases, the rotating wave approximation achieves the best performance when the driving is close to exact resonance.
On the other hand, in Fig.~\ref{fig:compare_full_reduced}(b), the frequency of the ac magnetic field is fixed at a detuned value $\omega=1.5Q$, while the driving strength of the ac magnetic field is varied.
The result shows that the performance of the rotating wave approximation becomes worse when the driving strength of the ac magnetic field increases.
In addition, it is obvious that the RWA applied in the full Hilbert space always performs better than the RWA applied in the reduced Hilbert space.
These results are in line with expectations since the conventional rotating wave approximation usually works well when the driving frequency is close to resonance and the driving strength is relatively weak.
These results also indicate that, when the driving strength is relatively strong or the driving frequency is off resonant, transitions outside the target resonance subspace or transitions between the target resonance subspace with the outside subspace cannot be neglected, and the rotating wave approximation should better be applied in the full Hilbert space.

\section{Counterrotating hybridized rotating wave method in the reduced Hilbert space\label{sc:chrw}}

For the standard rotating wave approximation, the counter-rotating terms in the Hamiltonian are neglected.
However, when the driving is strong or off-resonant, the effects induced by the counter-rotating terms appear and cannot be simply neglected.
In this section, we use the counterrotating hybridized rotating wave (CHRW) method introduced in Ref.~\cite{lv2012effects} to take into account the effects from these counter-rotating terms.
In order to make an analytic investigation, in this section, we will focus on the situation that the dynamics is restricted in the reduced Hilbert space.

\subsection{CHRW method for the effective three-level system}
We first consider the case that the reduced Hilbert space is spanned by $|I,M=1\rangle$, $|I,M=0\rangle$ and $|I,M=-1\rangle$, which can be regarded as an effective spin-1 system.
Corresponding to Eq.~(\ref{eq:reduced_spin_1}), the Hamiltonian of the system is
\begin{equation}
\label{eq:spin1_ac}
H(t)=QS_z^2+B_0S_z+B_1\cos(\omega t)S_x,
\end{equation}
where we have redefined $\sqrt{3}B_1\rightarrow B_1$ to make the analytic expression compact.
For Spin $S=1$, we have the following relation:
\begin{equation}
\label{eq:rwa_spin1}
e^{-iS_z^2\phi}S_xe^{iS_z^2\phi}=S_x\cos{\phi}+(S_yS_z+S_zS_y)\sin{\phi}.
\end{equation}

\begin{proof}
Suppose
\begin{equation}
e^{-iS_z^2\phi}S_xe^{iS_z^2\phi}=f(\phi),
\end{equation}
we have
\begin{equation}
\begin{aligned}
\frac{df}{d\phi}&=e^{-iS_z^2\phi}(-iS_z^2S_x+iS_xS_z^2)e^{iS_z^2\phi}\\
&=e^{-iS_z^2\phi}(S_yS_z+S_zS_y)e^{iS_z^2\phi},
\end{aligned}
\end{equation}
and
\begin{equation}
\frac{d^2f}{d\phi^2}=-e^{-iS_z^2\phi}S_xe^{iS_z^2\phi}=-f,
\end{equation}
so we have
\begin{equation}
\frac{d^2f}{d\phi^2}+f=0.
\end{equation}
The general solution is
\begin{equation}
f(\phi)=A\cos\phi+B\sin\phi,
\end{equation}
and we can obtain that
\begin{equation}
\begin{aligned}
A&=f(\phi=0)=S_x,\\
B&=f'(\phi=0)=S_yS_z+S_zS_y.
\end{aligned}
\end{equation}
\end{proof}

In order to apply the CHRW, firstly, we need to make the unitary transformation
\begin{equation}
U_1(t)=\exp{[-i\frac{B_1}{\omega}\xi\sin(\omega t)S_x]},
\end{equation}
where $\xi$ is an adjustable parameter.
The effective Hamiltonian after applying this unitary transformation can be obtained using
\begin{equation}
H_{\text{eff}}=U_1^\dagger(t)H(t)U_1(t)-iU_1^\dagger(t)\frac{d U_1(t)}{d t}.
\end{equation}
It is easy to check that,
\begin{eqnarray*}
U_1^\dagger(t)[B_1\cos(\omega t)S_x]U_1(t)&=&B_1\cos(\omega t)S_x,\\
-iU_1^\dagger(t)\frac{\partial U_1(t)}{\partial t}&=&-\xi B_1\cos(\omega t)S_x,
\end{eqnarray*}
and the transformation on the first two terms of Eq.~(\ref{eq:spin1_ac}) can be obtained as well, 
\begin{equation*}
\begin{aligned}
U_1^\dagger(t)(B_0S_z)U_1(t)=&B_0(\cos\varphi S_z+\sin\varphi S_y),\\
U_1^\dagger(t)(QS_z^2)U_1(t)=&Q\cos^2\varphi S_z^2+Q\sin^2\varphi S_y^2\\
&+Q\sin\varphi\cos\varphi(S_yS_z+S_zS_y),
\end{aligned}
\end{equation*}
where $\varphi=\frac{B_1}{\omega}\xi\sin(\omega t)$.
To proceed, we make an approximation by utilizing the identity
\begin{equation}
\exp{(iz\sin\alpha)}=\sum_{n=-\infty}^\infty J_n(z)e^{in\alpha},
\end{equation}
where $J_n(z)$ is the $n$th-order Bessel function of the first kind.
This approximation aims to neglect those high order harmonic terms, namely,
\begin{equation}
\exp{(iz\sin\alpha)}\approx J_0(z)+J_1(z)e^{i\alpha}+J_{-1}(z)e^{-i\alpha},
\end{equation}
so we have
\begin{equation*}
\begin{aligned}
\cos(2\varphi)
=&\frac{e^{i\frac{2B_1}{\omega}\xi\sin(\omega t)}+e^{-i\frac{2B_1}{\omega}\xi\sin(\omega t)}}{2}\\
=&\frac{1}{2}[J_0(\frac{2B_1}{\omega}\xi)+J_1(\frac{2B_1}{\omega}\xi)e^{i\omega t}+J_{-1}(\frac{2B_1}{\omega}\xi)e^{-i\omega t}\\
&+J_0(\frac{2B_1}{\omega}\xi)+J_1(\frac{2B_1}{\omega}\xi)e^{-i\omega t}+J_{-1}(\frac{2B_1}{\omega}\xi)e^{i\omega t}]\\
=&J_0(\frac{2B_1}{\omega}\xi),
\end{aligned}
\end{equation*}
and
\begin{equation*}
\begin{aligned}
\sin(2\varphi)
=&\frac{e^{i\frac{2B_1}{\omega}\xi\sin(\omega t)}-e^{-i\frac{2B_1}{\omega}\xi\sin(\omega t)}}{2i}\\
=&\frac{1}{2i}[J_0(\frac{2B_1}{\omega}\xi)+J_1(\frac{2B_1}{\omega}\xi)e^{i\omega t}+J_{-1}(\frac{2B_1}{\omega}\xi)e^{-i\omega t}\\
&-J_0(\frac{2B_1}{\omega}\xi)-J_1(\frac{2B_1}{\omega}\xi)e^{-i\omega t}-J_{-1}(\frac{2B_1}{\omega}\xi)e^{i\omega t}]\\
=&2J_1(\frac{2B_1}{\omega}\xi)\sin(\omega t),
\end{aligned}
\end{equation*}
where we have used the property that $J_{-n}(z)=(-1)^nJ_n(z)$.
Finally, we obtain the effective Hamiltonian after applying the unitary transformation $U_1(t)$ as follows:
\begin{equation}
\begin{aligned}
H_{\text{eff}}=&\frac{1+J_0(\frac{2B_1}{\omega}\xi)}{2}QS_z^2+\frac{1-J_0(\frac{2B_1}{\omega}\xi)}{2}QS_y^2\\
&+J_1(\frac{2B_1}{\omega}\xi)Q\sin(\omega t)(S_yS_z+S_zS_y)\\
&+B_1(1-\xi)\cos(\omega t)S_x+B_0J_0(\frac{B_1}{\omega}\xi)S_z\\
&+2B_0J_1(\frac{B_1}{\omega}\xi)\sin(\omega t)S_y.\\
\end{aligned}
\end{equation}

In order to apply the rotating reference frame transformation defined by Eq.~(\ref{eq:rwa_spin1}), we need to determine the value of $\xi$, which satisfies
\begin{equation}
J_1(\frac{2B_1}{\omega}\xi)Q=B_1(1-\xi)\equiv B_1^\prime,
\end{equation}
thus the effective Hamiltonian becomes
\begin{equation}
\begin{aligned}
H_{\text{eff}}=&\frac{1+J_0(\frac{2B_1}{\omega}\xi)}{2}QS_z^2+\frac{1-J_0(\frac{2B_1}{\omega}\xi)}{2}QS_y^2\\
&+B_1^{\prime}e^{-i\omega S_z^2 t}S_xe^{i\omega S_z^2 t}\\
&+B_0J_0(\frac{B_1}{\omega}\xi)S_z+2B_0J_1(\frac{B_1}{\omega}\xi)\sin(\omega t)S_y.\\
\end{aligned}
\end{equation}
To transform into the rotating reference frame, we make the unitary transformation
\begin{equation}
U_2(t)=e^{-i\omega S_z^2 t}
\end{equation}
and the effective Hamiltonian in the rotating frame becomes
\begin{equation}
\begin{aligned}
H_{\text{eff}}^{\prime}=&U_2^\dagger(t)H_{\text{eff}}U_2(t)-iU_2^\dagger(t)\frac{d U_2(t)}{d t}\\
=&[\frac{1+J_0(\frac{2B_1}{\omega}\xi)}{2}Q-\omega]S_z^2+\frac{1-J_0(\frac{2B_1}{\omega}\xi)}{2}QS_y^2\\
&+B_1^{\prime}S_x+B_0J_0(\frac{B_1}{\omega}\xi)S_z\\
&+2B_0J_1(\frac{B_1}{\omega}\xi)e^{i\omega S_z^2 t}\sin(\omega t)S_ye^{-i\omega S_z^2 t},
\end{aligned}
\end{equation}
where we have used the relation that $[S_z^2,S_y^2]=0$ for spin $S=1$.

In order to eliminate the time dependence of the last term in $H_{\text{eff}}^{\prime}$, we need to employ the standard rotating wave approximation, which is reasonable, since here we consider the situation that quadrupole interaction dominates, namely, $Q\gg B_0$.
For spin $S=1$, we also have the following relation:
\begin{equation}
\label{eq:rwa_spin1_2}
\begin{aligned}
2\sin(\omega t)S_y=&e^{-i\omega S_z^2 t}(S_xS_z+S_zS_x)e^{i\omega S_z^2 t}\\
&-e^{i\omega S_z^2 t}(S_xS_z+S_zS_x)e^{-i\omega S_z^2 t}.
\end{aligned}
\end{equation}
After applying the standard rotating wave approximation, the effective Hamiltonian becomes time-independent in the rotating reference frame,
\begin{equation}
\begin{aligned}
H_{\text{eff}}^{\prime}
\approx&\left[\frac{1+J_0(\frac{2B_1}{\omega}\xi)}{2}Q-\omega\right]S_z^2+\frac{1-J_0(\frac{2B_1}{\omega}\xi)}{2}QS_y^2\\
&+B_1^{\prime}S_x+B_0J_0(\frac{B_1}{\omega}\xi)S_z\\
&+B_0J_1(\frac{B_1}{\omega}\xi)(S_xS_z+S_zS_x).
\end{aligned}
\end{equation}

Eventually, the evolution operator corresponding to Eq.~(\ref{eq:spin1_ac}) after employing the CHRW method can be written as
\begin{equation}
\label{eq:spin1_chrw}
\mathcal{U}_3^\prime(t)=\exp{[-i\frac{B_1}{\omega}\xi\sin(\omega t)S_x]}\exp(-iwS_z^2t)\exp(-iH_{\text{eff}}^\prime t).
\end{equation}
Similarly, we can also obtain the explicit form of the evolution operator as in Eq.~(\ref{eq:Ut_explicit_spin1}), except now with $H_{\text{eff}}\rightarrow H_{\text{eff}}^{\prime}$ and with
\begin{equation}
\begin{aligned}
u=&2 B_0^2 \left[J_0(\frac{B_1}{\omega}\xi)^2+J_1(\frac{B_1}{\omega}\xi)^2\right]-2 B_1^2(1-\xi)^2\\
&+\frac{1}{2} \left[\omega -J_0(\frac{2B_1}{\omega}\xi) Q\right]^2+\frac{1}{6} (Q-\omega )^2,\\
\mathcal{R}=&\left[\cos\frac{B_1\xi\sin(\omega t)}{\omega}S_x^2-S_x^2+i\sin\frac{B_1\xi\sin(\omega t)}{\omega} S_x+\mathbb{1}_3\right]\\
&\times\left[(e^{-i\omega t}-1)S_z^2+\mathbb{1}_3\right].
\end{aligned}
\end{equation}
The overall evolution matrix for the whole Hilbert space has the same form as Eq.~(\ref{eq:spin1_reduced}), except with the change that $\mathcal{U}_3(t)\rightarrow\mathcal{U}_3^\prime(t)$.

\begin{figure}
\includegraphics[width=0.5\textwidth]{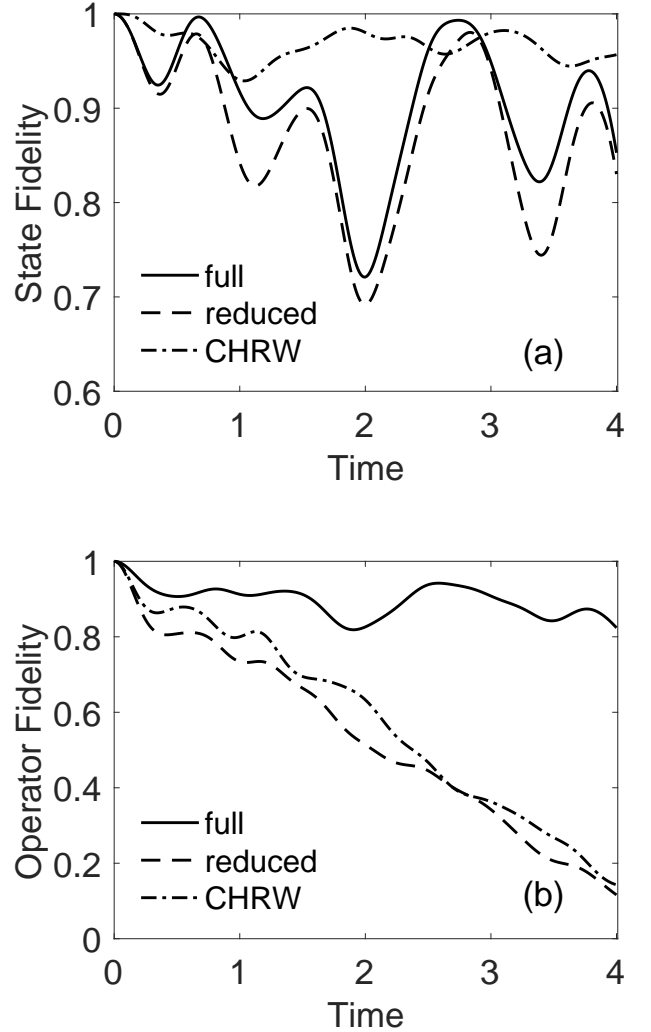}
\centering
\caption{\label{fig:on_resonant}Comparison of the state fidelity and the operator fidelity after employing the rotating wave approximation for spin $I=3$ with the on resonant ac magnetic field . The initial state of the system is $|\psi(0)\rangle=|I=3,M=0\rangle$, with the ac magnetic field tuned on resonant with the transition between $|I=3,M=\pm 1\rangle$ and $|I=3,M=0\rangle$, namely, $\omega=Q$. Other parameters are set as follows: $B_1=0.5Q$, $B_0=0.05Q$. The evolution time is in the units of $T_\pi$. (a) Comparison of the state fidelity after applying the rotating wave approximation in the full Hilbert space (solid line) and in the reduced Hilbert space (dashed line for the standard RWA and dash-dotted line for the CHRW method). (b) Comparison of the operator fidelity after applying the rotating wave approximation in the full Hilbert space (solid line) and in the reduced Hilbert space (dashed line for the standard RWA and dash-dotted line for the CHRW method).}
\end{figure}

\subsection{CHRW method for the effective two-level system}
The evolution operator corresponding to the reduced two-level system using the CHRW method can be obtained in a similar way.
We now briefly describe the procedure.
The Hamiltonian for the reduced Hilbert space equivalent to a two-level system is Eq.~(\ref{eq:two_level}), namely,
\begin{equation*}
H_1(t)=\frac{\omega_0}{2}\hat{\sigma}_z+B_1^\prime \cos(\omega t)\hat{\sigma}_x.
\end{equation*}
After making the first unitary transformation
\begin{equation}
U_1(t)=\exp{[-i\frac{B_1^\prime}{\omega}\xi\sin(\omega t)\hat{\sigma}_x]},
\end{equation}
the effective Hamiltonian becomes
\begin{equation}
\label{eq:two_level_chrw}
H_{\text{eff}}=\frac{\omega_0}{2}(\cos\varphi \hat{\sigma}_z+\sin\varphi \hat{\sigma}_y)+B_1^\prime (1-\xi)\cos(\omega t)\hat{\sigma}_x,
\end{equation}
where $\varphi=\frac{2B_1^\prime}{\omega}\xi\sin(\omega t)$.
Similar to previous discussions, after dropping those high order harmonic terms, we obtain
\begin{equation}
\begin{aligned}
\cos\varphi&=J_0(\frac{2B_1^\prime}{\omega}\xi),\\
\sin\varphi&=2J_1(\frac{2B_1^\prime}{\omega}\xi)\sin(\omega t).
\end{aligned}
\end{equation}
With the condition
\begin{equation}
\omega_0J_1(\frac{2B_1^\prime}{\omega}\xi)=B_1^\prime(1-\xi)\equiv B_{\text{eff}},
\end{equation}
Eq.~(\ref{eq:two_level_chrw}) becomes
\begin{equation}
H_{\text{eff}}=\frac{\omega_0}{2}J_0(\frac{2B_1^\prime}{\omega}\xi)\hat{\sigma}_z+B_{\text{eff}}e^{-i\omega \hat{\sigma}_zt/2}\hat{\sigma}_xe^{i\omega \hat{\sigma}_zt/2}.
\end{equation}
We then transform the Hamiltonian into the rotating reference frame by applying the second unitary transformation
\begin{equation}
U_2(t)=e^{-i\omega \hat{\sigma}_zt/2},
\end{equation}
and the effective Hamiltonian in the rotating frame now becomes time-independent,
\begin{equation}
H_{\text{eff}}^\prime=\left[\frac{\omega_0}{2}J_0(\frac{2B_1^\prime}{\omega}\xi)-\frac{\omega}{2}\right]\hat{\sigma}_z+B_{\text{eff}}\hat{\sigma}_x.
\end{equation}
Finally, the evolution matrix for this reduced two-dimensional Hilbert space can be written as
\begin{equation}
\mathcal{U}_1^\prime(t)=\exp{[-i\frac{B_1^\prime}{\omega}\xi\sin(\omega t)\hat{\sigma}_x]}\exp({-i\frac{\omega}{2} \hat{\sigma}_zt})\exp{(-iH_{\text{eff}}^\prime t)}.
\end{equation}
We can obtain $\mathcal{U}_2^\prime(t)$ using the same procedure and the overall evolution matrix for the whole Hilbert space has the same form as Eq.~(\ref{eq:Ut_two_level}), except with the change that $\mathcal{U}_1(t)\rightarrow \mathcal{U}_1^\prime(t)$ and $\mathcal{U}_2(t)\rightarrow \mathcal{U}_2^\prime(t)$.

\section{Comparison of the state fidelity and the operator fidelity\label{sc:comparison}}

\begin{figure}
\includegraphics[width=0.5\textwidth]{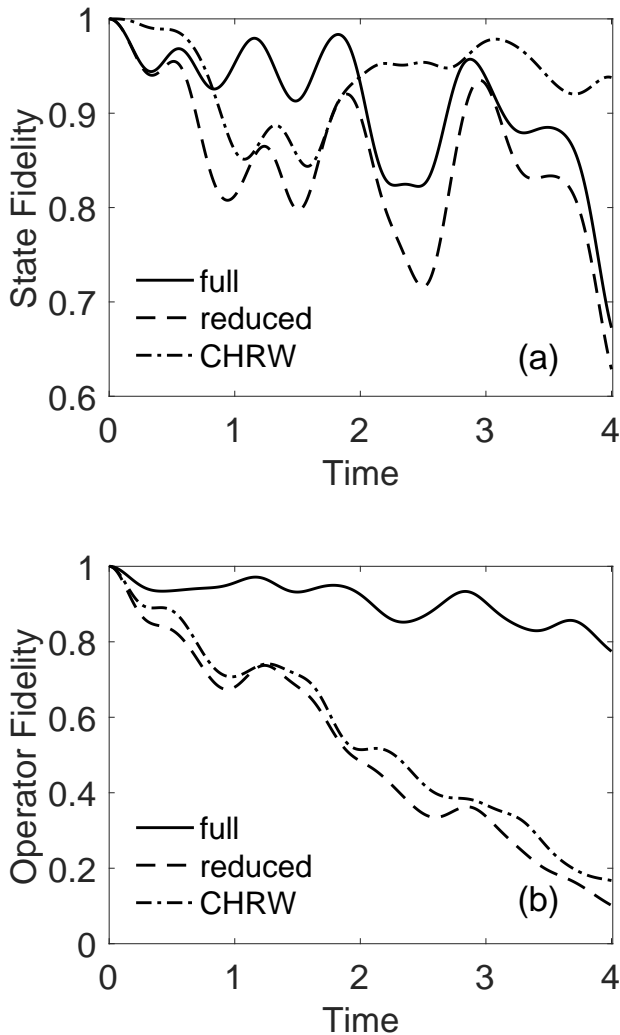}
\centering
\caption{\label{fig:off_resonant}Comparison of the state fidelity and the operator fidelity after applying the rotating wave approximation for spin $I=3$ with the near resonant ac magnetic field. The parameters used here are the same as in Fig.~\ref{fig:on_resonant}, except that the ac magnetic field is tuned off resonant, $\omega=1.5Q$.}
\end{figure}

%

In order to compare the performance of the evolution operator obtained after employing the rotating wave approximation in the reduced Hilbert space or in the full Hilbert space, we calculate the quantum state fidelity and the quantum operator fidelity.
The operator fidelity is defined as~\cite{wang2009operator}
\begin{equation}
\label{eq:operator_fidelity}
F=\left|\frac{1}{N}\text{Tr}\left[\mathcal{U}^\dagger(t)U(t) \right]\right|^2,
\end{equation}
where $\mathcal{U}(t)$ is the evolution operator obtained after applying the rotating wave approximation, while $U(t)$ is the propagator obtained by exactly solving the Schr\"{o}dinger equation
\begin{equation}
i\frac{\partial U(t)}{\partial t}=H(t)U(t),
\end{equation}
using the Runge-Kutta method with the initial condition that $U(0)=\mathbb{1}$.
On the other hand, the state fidelity is defined as
\begin{equation}
f=\sqrt{|\langle \psi(t)|\psi^\prime(t)\rangle|^2},
\end{equation}
where $|\psi(t)\rangle=U(t)|\psi(0)\rangle$ and $|\psi^\prime(t)\rangle=\mathcal{U}(t)|\psi(0)\rangle$.

In Figs.~\ref{fig:on_resonant} and~\ref{fig:off_resonant}, we compare the state fidelity and the operator fidelity after applying the rotating wave approximation in the full Hilbert space or in the reduced Hilbert space for spin $I=3$.
The initial state of the system is prepared into $|\psi(0)\rangle=|I=3,M=0\rangle$.
The ac magnetic field is tuned on resonant (Fig.~\ref{fig:on_resonant}) or near resonant (Fig.~\ref{fig:off_resonant}) with the transition between $|I=3,M=\pm 1\rangle$ and $|I=3,M=0\rangle$, so here the reduced Hilbert space is equivalent to a spin-1 system and the evolution operator in the reduced Hilbert space corresponds to Eq.~(\ref{eq:Ut_explicit_spin1}) for the standard RWA, or Eq.~(\ref{eq:spin1_chrw}) for the CHRW method.
Meanwhile, the evolution operator after employing the rotating wave approximation in the full Hilbert space corresponds to Eq.~(\ref{eq:Ut_full}).
The characteristic time for the time evolution is defined as the the time needed for a $\pi$ rotation when using the standard RWA in the reduced subspace, which is
\begin{equation}
\label{eq:t_pi}
T_{\pi}=\frac{\pi}{\sqrt{\frac{1}{2}I(I+1)B_1^2+(Q-\omega)^2}}.
\end{equation}
Here, we mainly consider the situation that the driving strength is relatively strong, with $B_1=0.5Q$.

The result presented in Fig.~\ref{fig:on_resonant} shows that, when the driving is on resonant with the target transition, the CHRW method in the reduced Hilbert space exhibits the best state fidelity, while the standard RWA in the full Hilbert space performs better than the standard RWA in the reduced Hilbert space.
However, when the operator fidelity is compared, the standard RWA in the full Hilbert space achieves the best fidelity, while the CHRW method performs better than the standard RWA in the reduced Hilbert space.
It is essential to note that the operator fidelity does not depend on the initial state, while the state fidelity relies on the specific initial state of the system.
The result presented in Fig.~\ref{fig:off_resonant}, corresponding to the off resonant driving, exhibits similar features as the on resonant case, except that the advantage on the state fidelity of the CHRW method in the reduced Hilbert space becomes less significant than the standard RWA in the full Hilbert space.

Basically, there are two factors affecting the performance of the rotating wave approximation applied in the high-spin system with quadrupole interaction: the effects of the counter-rotating terms and the leakage from the target resonance subspace.
The CHRW method takes into account the effects of the counter-rotating terms, while the RWA in the full Hilbert space takes into account the leakage from the target resonance subspace, so both of them perform better than the standard RWA in the reduced Hilbert space, especially when the driving strength is relatively strong.
On the other hand, when operator fidelity of these rotating wave approximation methods is estimated, the RWA in the full Hilbert space always performs better than the RWA/CHRW in the reduced Hilbert space.
Since the operator fidelity does not depend on the initial state of the system, when this initial state has components outside the target resonance subspace, the ac magnetic field will unavoidably result in transitions among these components, or lead to transitions between these components and the target resonance subspace, whereas the treatment using the reduced Hilbert space simply neglects these transitions.

The above results indicate that, when the driving is off resonant or the driving strength is relatively strong, the leakage from the target resonance subspace cannot be neglected anymore, and the employment of the RWA in the full Hilbert space becomes essential.
Nevertheless, for some circumstances, like when the evolution of the quantum state can be approximately restricted in the target resonance subspace, the CHRW method can be utilized.
On the other hand, to estimate the performance of quantum processes in the high-spin system with quadrupole interaction, the RWA in the full Hilbert space should be preferred.


\section{Discussion\label{sc:discussion}}
Naturally, the application of the CHRW method in the full Hilbert space will give much higher operator fidelity and state fidelity, however, due to the lack of identities like Eq.~(\ref{eq:rwa_spin1}) and Eq.~(\ref{eq:rwa_spin1_2}), the analytic investigation on this combination for spin $I>1$ is rather challenging.
Maybe alternative methods will be found to apply the CHRW method in the full Hilbert space in the future.
Although we only considered the application of classical fields in this paper, the theory presented here may be generalized to the case with quantum fields as well.

The theory proposed in this paper can be used to construct high-fidelity quantum gate for quantum computing~\cite{nielsen2002quantum} using quantum spin systems with quadrupole interaction presents, especially for the situation with strong driving~\cite{fuchs2009gigahertz}.
Similarly, the theory can be utilized to enhance the performance of quantum sensing~\cite{degen2017quantum} or quantum metrology~\cite{Giovannetti2011advances} using high-spin quantum spin systems.
Besides, due to the similarity of the Hamiltonian investigated in this paper with the spin squeezing Hamiltonian~\cite{ma2011quantum}, some results in this paper may find applications in quantum spin squeezing.

%
\begin{acknowledgments}
This work was supported by the National Key Research and Development Program of China (Grants No. 2017YFA0304202 and No. 2017YFA0205700), the NSFC through Grant No. 11875231 and No. 11935012, and the Fundamental Research Funds for the Central Universities through Grant No. 2018FZA3005.
\end{acknowledgments}

\appendix


\bibliographystyle{apsrev4-1}
%

\end{document}